\documentclass[onecolumn,showpacs]{revtex4}

\usepackage{graphicx}
\usepackage{dcolumn}
\usepackage{bm}

\input epsf

\begin{document}

\title{\Large  BLACK HOLES IN THE EINSTEIN-GAUSS-BONNET THEORY AND THE GEOMETRY OF THEIR
THERMODYNAMICS-$II$ }

\author{\bf Ritabrata
Biswas\footnote{ritabratabiswas@rediffmail.com}~And~Subenoy
Chakraborty.\footnote{schakraborty@math.jdvu.ac.in} }

\affiliation{Department Of Mathematics, Jadavpur University}

\date{\today}

\begin{abstract}

In the present work we study $(i)$ charged black hole in
Einstein-Gauss-Bonnet (EGB) theory, known as
Einstein-Maxwell-Gauss-Bonnet (EMGB) black hole and $(ii)$ black
hole in EGB gravity with Yang-Mills field. The thermodynamic
geometry of these two black hole solutions has been investigated,
using the modified entropy in Gauss-Bonnet theory.

\end{abstract}

\pacs{95.36.+x,04.70.Dy, 04.60.Kz.}

\maketitle

\section{INTRODUCTION}

The interest in the black hole thermodynamics results due to the
nice similarities between a black hole and a thermodynamical
system. Also the thermodynamical quantities, namely the
temperature (known as Hawking temperature) and entropy of a black
hole are related to the geometry of the event horizon: the
temperature is proportional to the surface gravity of the event
horizon while the entropy is related to the area of the event
horizon $[1,2]$ and they satisfy the first law of thermodynamics
$[3]$. However, it is still a challenging problem to find any
statistical origin of the black hole thermodynamics.

In this context, Ruppeiner $[4]$ introduced a metric as the
Hessian matrix of the thermodynamic entropy (known as Ruppeiner
metric) and showed that the corresponding geometry has physical
relevance in the fluctuation theory of equilibrium thermodynamics.
More precisely, the Ruppeiner geometry is related to the phase
structure of thermodynamic system : the scalar curvature $(R)$ of
the Ruppeiner metric diverges (i.e. $R \rightarrow -\infty$) at
the phase transition and critical point while $R \rightarrow 0$
(i.e. flat Ruppeiner metric) indicates no statistical interaction
of the thermodynamical system. Also the inverse Ruppeiner metric
gives the second moments of fluctuations. It should be mentioned
in this context that Weinhold $[5]$ first introduced the geometric
concept into ordinary thermodynamics by introducing a Riemannian
metric (known as Weinhold metric) as the Hessian of the internal
energy(mass parameter here), having no phyical meaning in
equilibrium thermodynamics. However, the Ruppeiner metric is
conformally related to the Weinhold metric. In this paper we deal
with five dimensional black hole solutions in $(i)$
Einstein-Maxwell-Gauss-Bonnet (EMGB) theory with a cosmological
constant and $(ii)$ Einstein-Yang-Mills-Gauss-Bonnet (EYMGB)
theory and the corresponding thermodynamics with modified form of
entropy $[6]$. The entropy is modified by the entropy of the
analogous Schwarzchild black hole solution in EGB theory. The
black hole solution in EMGB theory and the geometry of its
thermodynamic has been studied in section $II$ while section $III$
contains the black hole solution in EYMGB theory and the
corresponding thermodynamic geometry.

\section{GEOMETRIC IDEA OF 5-D EMGB BLACK HOLE THERMODYNAMICS}

We present the black hole solution and its properties in five
dimensional Einstein-Maxwell-Gauss-Bonnet theory and subsequently
the thermodynamics of this black hole is studied.

A five dimensional spherically symmetric solution in
Einstein-Maxwell theory with Gauss-Bonnet term was obtained by
Wiltshire $[7]$ with metric ansatz
\begin{equation}
ds^{2}=-B(r)dt^{2}+\frac{dr^{2}}{B(r)}+r^{2}d\Omega_{3}^{2}
\end{equation}

where,
\begin{equation}
B(r)=1+\frac{r^{2}}{4\alpha}-\frac{r^{2}}{4\alpha}\sqrt{1+\frac{16
m \alpha}{\pi
r^{4}}-\frac{8q^{2}\alpha}{3r^{6}}+\frac{4\Lambda\alpha}{3}}
\end{equation}

and $d\Omega_{3}^{2}$  is the metric of unit three sphere.

The electric field is chosen along the radial direction having non
vanishing components of the electromagnetic tensor $F_{\mu \nu}$
(in an orthonormal frame) are $$ F_{\hat{t}\hat{r}}=
-F_{\hat{r}\hat{t}} =\frac{q}{4 \pi r^{3}}$$.

The two parameters $m(>0)$ and $q$ are identified as the mass and
charge of the system. In the limit $\alpha \rightarrow 0$ the
above solution reduces to the five dimensional
Reissner-Nordstr$\ddot{o}$m solution with a cosmological constant.
Note that the solution $(2)$ is well defined if the expression
within the square root is positive definite. As a result the
solution is valid for $r>r_{0}$ where $r_{0}$ is the largest
positive root of the equation $[7-9]$
\begin{equation}
\left(3+4\Lambda \alpha\right)r^{6}+\left(\frac{48 m
\alpha}{\pi}\right)r^{2}-8 \alpha q^{2}=0
\end{equation}

This hyper surface $r=r_{0}$ corresponds to curvature singularity
which may be covered by the event horizon (for black hole
solution) or may represent a naked singularity.

If $r_{h}$ is the radius of the event horizon then it is related
to the mass $(m)$ and charge $(q)$ of the black hole by the
relation
\begin{equation}
\frac{\Lambda}{3}r_{h}^{6}-2r_{h}^{4}+\left(\frac{4m}{\pi}-4
\alpha\right)r_{h}^{2}-\frac{2 q^{2}}{3}=0
\end{equation}

(Note that event horizon exits if the above equation has at least
one positive root)

So we can write the mass parameter as
\begin{equation}
m=\pi \alpha+\frac{\pi
q^{2}}{6}r_{h}^{-2}+\frac{\pi}{2}r_{h}^{2}-\frac{\pi
\Lambda}{12}r_{h}^{4}
\end{equation}
Now using the entropy of the horizon $[6]$ of the of the
spherically symmetric black hole solution (Schwarzhild solution)
in Einstein-Gauss-Bonnet theory, the entropy of the present black
hole takes the form (choosing the Boltzmann constant
appropriately)
\begin{equation}
S=r_{h}^{3}+6 \tilde{\alpha} r_{h}
\end{equation}
where, $\tilde{\alpha}=(n-2)(n-3)\alpha$ and $\alpha$ is the
Gauss-Bonnet coupling parameter having dimension $(length)^{2}$.
From equation $(5)$ and $(6)$ we can say that the mass parameter
$(m)$ is in principle a function of entropy $(S)$ and charge
$(q)$. Using the energy conservation law of the black hole (i.e.
$dm=TdS+\phi dq$) one obtains the temperature and electric
potential of the black hole on the event horizon as
\begin{equation}
T=\left(\frac{\partial m}{\partial S}\right)_{q}=\frac{\pi
\left(3r_{h}^{4}-\Lambda
r_{h}^{6}-q^{2}\right)}{9r_{h}^{3}\left(r_{h}^{2}+2\tilde{\alpha}\right)}
\end{equation}
and
\begin{equation}
\phi=\left(\frac{\partial m}{\partial q}\right)_{S}=\frac{\pi
q}{3}r_{h}^{-2}
\end{equation}

The Weinhold metric $[5, 10]$, the Hessian of the mass parameter,
i.e.,
\begin{equation}
g_{i j}^{(W)} = \partial_{i}\partial_{j}m(S,q)
\end{equation}
has the explicit form
\begin{equation}
ds_{W}^{2} = \frac{\pi}{9
r_{h}^{3}\left(r_{h}^{2}+2\tilde{\alpha}\right)}\left[\frac{U(r_{h})}{3r_{h}\left(r_{h}^{2}+
2\tilde{\alpha}\right)^{2}}dS^{2}-4qdSdq+3r_{h}\left(r_{h}^{2}+2\tilde{\alpha}\right)dq^{2}\right]
\end{equation}
with,
$$U(r_{h}, q)=-\Lambda r_{h}^{8}-3r_{h}^{6}\left(1+2\tilde{\alpha} \Lambda\right)+6\tilde{\alpha}r_{h}^{4}
+5q^{2}r_{h}^{2}+6\tilde{\alpha}q^{2}$$

The transformation,
\begin{equation}
S=r_{h}^{3}+6\tilde{\alpha}r_{h}~,~x=r_{h}~,~y=qr_{h}^{-2}
\end{equation}
reduces the above metric into the diagonal form as,
\begin{equation}
ds_{W}^{2}=\frac{\pi}{2}\left[-\left\{4y^{2}-\frac{U_{0}(x,
y)}{x^{4}\left(x^{2}+2\tilde{\alpha}\right)}\right\}dx^{2}+x^{2}dy^{2}\right]
\end{equation}
with
$$U_{0}(x,y)=-\Lambda x^{8}-3x^{6}\left(1+2\tilde{\alpha}\Lambda\right)+6\tilde{\alpha}x^{4}+5y^{2}x^{6}
+6\tilde{\alpha}y^{2}x^{4}$$

As the Ruppeiner metric, the Hessian of the entropy function,
i.e.,
$$g_{i j}^{(R)} = \partial_{i}\partial_{j}S(m,q)$$

is conformally related to the Weinhold metric by the conformal
factor $T^{-1}$ so we have,
$$ds_{R}^{2}=\frac{1}{T}ds_{W}^{2}$$

Thus the explicit form of Ruppeiner metric is
\begin{equation}
ds_{R}^{2}=\frac{3\left(x^{2}+2\tilde{\alpha}\right)}{x\left(3-\Lambda
x^{2}-y^{2}\right)}\left[-\left\{4y^{2}-\frac{U_{0}(x,
y)}{x^{4}\left(x^{2}+2\tilde{\alpha}\right)}\right\}dx^{2}+x^{2}dy^{2}\right]
\end{equation}

The non-flat nature of the Ruppeiner metric suggests that the
statistical mechanics description is possible for the
thermodynamics of the present black hole. As the expression for
scalar curvature $(R)$ is very complicated so we have not
presented it here but one thing to note that $R$ diverges at
$y=\pm\sqrt{3-\Lambda x ^{2}}$.

Further for a given charge, the expression for the heat capacity
is
\begin{equation}
c_{q}=\frac{3x^{5}\left(x^{2}+2\tilde{\alpha}\right)^{2}\left(3-\Lambda
x^{2}-y^{2}\right)}{U_{0}(x, y)}
\end{equation}

Thus at $y^{2}=3-\Lambda x^{2}$, the metric coefficients as well
as the curvature scalar becomes singular while $c_{q}$ changes
sign at that point. So there is a phase transition $[11]$
corresponding to $r_{h}$ given by $$q^{2}+\Lambda r_{h}^{6}-3
r_{h}^{4}=0$$.

Further, note that for positive $\Lambda$ , $U_{0}$ has one zero
at some positive x, so in some part of state space $c_{q}$ will be
positive , i.e., black hole will be a stable one while other part
corresponds to unstable black hole.

\section{THERMODYNAMICS OF 5-D EYMGB BLACK HOLE FROM GEOMETRIC ASPECT}

In EYMGB gravity theory, the 5-D spherically symmetric solution
obtained recently by Mazharimousavi and Halisoy $[12]$ has the
metric ansatz
\begin{equation}
ds^{2} = -U(r)dt^{2}+\frac{dr^{2}}{U(r)}+r^{2}d\Omega_{3}^{2}
\end{equation}
where
\begin{equation}
U(r)=1+\frac{r^{2}}{4 \alpha}\pm \sqrt{\left(\frac{r^{2}}{4
\alpha}\right)^{2}+\left(1+\frac{m}{2\alpha}\right)+\frac{q^{2}\ln
r}{\alpha}},
\end{equation}
$m$ is a constant of integration and q is the only non-zero gauge
charge such that
$$F^{i}_{\alpha \beta}F^{i \alpha \beta}=\frac{6q^{2}}{r^{4}}$$

with $F^{i}_{\alpha \beta}F^{i \alpha \beta}$, the Yang-Mills
field 2-form.

In the limit $\alpha\rightarrow 0$, we obtain Einstein-Yang-Mills
solution with $m$ as a mass of the system provided negative branch
of the above solution is chosen. Note that if the Gauss-Bonnet
coupling parameter $\alpha$ is positive definite then the solution
is well defined for all $r$ but for $\alpha<0$ , the geometry has
a curvature singularity at the hyper surface $r=r_{s}$ where,
$r_{s}$ is the largest value of the radial coordinate such that
the expression within the square root is positive. If $r_{h}$ is
the radius of the event horizon($r_{h}$ is the positive root of
the equation $U(r)=0$, the least one if there are more than one
positive roots) then black hole will exit if $r_{s}<r_{h}$,
otherwise the singularity will be naked.

Now putting$U(r)=0$, the event horizon satisfies
\begin{equation}
r_{h}^{2}-m-2q^{2} \ln r_{h}=0
\end{equation}

which is independent of the coupling parameter $\alpha$. Now using
the entropy for the Schwarzchild solution in EGB gravity, we have
(choosing the unit properly) as before
\begin{equation}
S=r_{h}^{3}+6\alpha r_{h}
\end{equation}

Thus the expressions for the thermodynamic quantities namely
temperature, electric potential, heat capacity are given by
$$T=\frac{2\left(r_{h}^{2}-q^{2}\right)}{3r_{h}\left(r_{h}^{2}+2\tilde{\alpha}\right)}$$
$$\phi= - 4 q \ln r_{h}$$
\begin{equation}
c_{q}=\frac{3r_{h}\left(r_{h}^{2}-q^{2}\right)\left(r_{h}^{2}+2\tilde{\alpha}\right)^{2}}{P(r_{h},q)}
\end{equation}
with
$$P(r_{h},q)=-r_{h}^{4}+r_{h}^{2}(2\tilde{\alpha}+3q^{2})+2q^{2}\tilde{\alpha}$$
Now the explicit form of the Weinhold metric is
\begin{equation}
ds_{W}^{2}=\left[\frac{2P(r_{h})}{9r_{h}^{2}\left(r_{h}^{2}+2\tilde{\alpha}\right)^{3}}dS^{2}
-\frac{8q}{3r_{h}\left(r_{h}^{2}+2\tilde{\alpha}\right)}dSdq-4\ln
r{h}dq^{2}\right]
\end{equation}
The transformation
\begin{equation}
S=r_{h}^{3}+6\tilde{\alpha}r_{h}~,~x=\ln r_{h}~,~y=q.x
\end{equation}
 makes the Weinhold metric and hence the Ruppeiner metric (conformally related to the Weinhold
 metric)into diagonal form as
\begin{equation}
ds_{R}^{2}=\frac{3e^{x}\left(e^{2 x}+2
\tilde{\alpha}\right)}{\left(e^{2x}-\frac{y^{2}}{x^{2}}\right)}\left[\left\{\frac{p_{0}(x,
y)}{e^{2x}\left(e^{2x}+2\tilde{\alpha}\right)}+2\frac{q^{2}}{x}\right\}dx^{2}-2\frac{dy^{2}}{x}\right]
\end{equation}
with $$p_{0}(x,
y)=-e^{4x}+e^{2x}\left(2\tilde{\alpha}+\frac{3y^{2}}{x^{2}}\right)+2\tilde{\alpha}\frac{y^{2}}{x^{2}}$$.
As the curvature scalar (that we have not presented here due to
its complicated and lengthy form) as well as the metric
coefficients have singularity at $y^{2}=x^{2}e^{2x}$ and the heat
capacity changes sign in crossing the above singularity so there
is a phase transition at the singularity. Further, the expression
for $c_{q}$ shows that it has singularities at
$$r_{h}^{4}- \left(2\tilde{\alpha}+3q^{2}\right)r_{h}^{2}-2q^{2}\tilde{\alpha}=0$$
i.e., at
$$r_{h}^{2}=\left(\tilde{\alpha}+\frac{3}{2}q^{2}\right)+\sqrt{\left(\tilde{\alpha}+\frac{3}{2}q^{2}\right)^{2}
+2q^{2}\tilde{\alpha}}$$. Thus $c_{q}$ is not positive definite so
the black hole is not a stable one in the whole admissible state
space.

For both the black holes the variation of the thermodynamical parameters (namely mass, Temperature and heat capacity)
are different from those of our earlier work $[13]$. The basic difference in the two papers is
 the choice of the entropy function in the previous paper;
  Bekentein-Hawking entropy relation has been used while in the present paper we have chosen the entropy
  of a Schwarzschild like black hole in EGB gravity. In fact, heat capacity shows a significant variation and as a
  result we may conclude that choice of entropy has an important role for the stability as well as phase transition
  of a black hole. \\\\\\

{\bf Acknowledgement:}\\

A part of the work is done during a visit to IUCAA. The authors
are thankful to IUCAA for warm hospitality and
facilities of research.\\\\\\\\

$[1]$ S. W. Hawking , {\it Commun. Math. Phys.} {\bf43}, 199(1975).\\

$[2]$ J. D. Bekenstein, {\it Phys. Rev. D} {\bf 7}, 2333(1973).\\

$[3]$ J. M. Bardeen, B. Carter and S. W. Hawking, {\it Commun.
Math. Phys.} {\bf31}, 161(1973).\\

$[4]$ G. Ruppeiner, {\it Phys Rev. A} {\bf 20}, 1608(1979).

~~~~~~~For details see

~~~~~G. Ruppeiner, {\it Rev. Mod. Phys.} {\bf67}, 605(1995);
{\bf68}, 313(E)(1996)\\

$[5]$ F. Weinhold, {\it J. Chem. Phy.} {\bf63}, 2479(1975).\\

$[6]$ R. C. Myers and J. Z. Simon, {\it Phys. Rev.D} {\bf 38},
2434 (1988); R. G. Cai, {\it Phys. Rev.D} {\bf65}, 084014 (2002);

~~~~~ R. G. Cai and Q. Guo {\it Phys. Rev.D} {\bf69}, 104025
(2004);

~~~~~T. Clunan, S. F. Ross and D.J. Smith{\it Class. Quant. Grav.}
{\bf21}, 3447(2004).

$[7]$ D. L. Wiltshire, {\it Phys. Letts. B.} {\bf 169}, 36(1986);
{\it Phys. Rev. D} {\bf 38}, 2445(1988). \\

$[8]$ D. G. Boulware and S. Deser, {\it Phys. Rev. Lett.} {\bf
55},2656(1985).\\

$[9]$ M. Thibeault, C. Simeone and E. F. Eirod, {\it Gen. Relt.
Grav.} {\bf38}, 1593(2006). \\

$[10]$ S. Chakraborty and T. Bandyopadhyay, {\it Class. Quant.
Grav.} {\bf25}, 245015(2008).

$[11]$ S. H. Mazharimousavi and M. Halisoy, {\it Phys. Rev.D} {\bf
76}, 087501(2007).\\

$[12]$ J. E. Aman and N. Pidokrajt, {\it Phys. Rev.D} {\bf 73},
024017(2006). J. E. Aman, I. Bengtsson and N. Pidokrajt {\it Gen.
Relt. Grav} {\bf 35}, 1733(2003).
\\

$[13]$ R. Biswas and S. Chakraborty, {\it arXiv-0905.1776(gr-qc)} (2009)\\

\end{document}